\documentclass[twocolumn]{aastex631}
\usepackage{amsmath}

\newcommand\numberthis{\addtocounter{equation}{1}\tag{\theequation}}

\begin{document}

\title{Tidal Dissipation from Circularization in Kepler Binaries}

\author[0000-0001-9241-5921]{Joshua Schussler}
\affiliation{The University of Texas at Dallas \\
800 W Campbell Rd \\
Richardson, Texas, USA}

\author[0009-0007-1309-3276]{Torsha Majumder}
\affiliation{The University of Lethbridge \\
4401 University Drive \\
Lethbridge, Alberta, Canada}

\author[0000-0003-4464-1371]{Kaloyan Penev}
\affiliation{The University of Texas at Dallas \\
800 W Campbell Rd \\
Richardson, Texas, USA}

\begin{abstract}

Understanding tidal dissipation is a requisite step for explaining the evolution of systems such as short-period binaries. This understanding has not yet been achieved, and in fact there are many different approaches to modelling tides. By using Bayesian analysis on a system-by-system basis, we provide additional constraints on dissipation for 105 short-period Sunlike \textit{Kepler} binaries. We account for period-dependent $Q$ and stellar evolution, and propagate observational uncertainties to uncertainties in our constraints.
We find a group constraint of $\log Q_*' \approx 6.75$, with no apparent dependence on tidal period. We also do not detect mass dependence for $Q$.
Our inferred prescription for the tidal dissipation successfully reproduces the unexpected overlap between almost perfectly circular and significantly non-circular orbits in binaries of Sunlike stars reported by J. J. Zanazzi. 

\end{abstract}


\section{Introduction} \label{sec:intro}

Tidal dissipation is a process that dissipates energy in an orbit and rotation as heat, playing a crucial role in binaries and star-planet systems. However, the mechanisms underlying tidal dissipation are poorly understood. There are many proposed models, but to test their predictions, observational data must be analyzed to determine real-life values—a bottleneck that impedes progress in the field. In this work, we have notably improved on existing analysis approaches in the literature; we present constraints on tidal dissipation for a large number of binary systems.

Tidal dissipation is most notable in interactions between objects that are very near to each other.
For example, a hot Jupiter—that is, a gas giant planet with a short period, circular orbit—may be subject to destruction via this mechanism \citep{destruction1,destruction2,destruction3}.
In binary star systems, the effects of tidal dissipation become noticeable at periods of around 30 days, with increasing strength as the period shrinks. This is also affected by factors like the eccentricity of the orbit and the size and mass of each star.
This work examines such short-period binaries, taking advantage of the unprecedented precision provided by the Kepler space telescope.

Different models offer different predictions for how properties like eccentricity, mass, or radius impact tidal dissipation. Broadly, tidal dissipation models can be split into two major categories: equilibrium and dynamical.

Equilibrium models assume that the primary source of dissipation is a quasi-static, but still time-dependent, tidal distortion. The main dissipative process arises from the interaction of the tidal distortion with the convective motion.
Despite this commonality, predictions can vary widely. Some examples can be found in \citet{old2}, \citet{iDontHaveTideForYourGamesQ}, \citet{1989A&A...223..112Z}, and \citet{starStar3}. 
The assumption dynamical tides models make, meanwhile, is that the time-dependent corrections to the steady-state are the main cause of dissipation, via interactions between tidal perturbations and internal waves. The exact kind of interaction depends on the model.
Examples of dynamical tides include \citet{ogilvie2007} and \citet{Barker_20}.

It is common to parameterize the rate of dissipation with the tidal quality factor, $Q$. The inverse of $Q$ is equal to the amount of energy in a tidal wave dissipated per radian that wave travels. This is often used in a slightly modified form:
\begin{equation}
    Q' \equiv Q / k_2.
\end{equation}
The tidal Love number $k_2$ is the real part of the ratio of the quadrupolar component of the induced gravitational potential of the tidally distorted object to the quadrupolar component of the external potential which is causing the deformation; generally the hydrostatic value is used.

Certain simplifications are frequently made in attempts to constrain tidal dissipation. For example, stellar evolution may not be taken into account, or the radius or structure of the orbiting bodies themselves could be held static. While $Q'$ is expected to have dependence on factors like tidal frequency, tidal amplitude, spin, and the internal structure of the dissipating body, these dependencies are sometimes ignored.

Our group's method of analysis avoids these issues. We focus on Sunlike stars, both as binaries and in star-planet systems.
Our orbital evolutions incorporate stellar evolution, and we use Markov Chain Monte Carlo (MCMC) to allow the uncertainties to inform the constraints on $Q$.
Furthermore, our treatment of $Q$ is a broken power law as a function of period, with the exponent as one of the MCMC variables.
We have applied this analysis to open cluster binaries (\citet{itme}, hereafter PS22) and hot Jupiter systems \citep{itme3}, as well as Kepler binaries with a focus on spin evolution (\citet{itme2}, hereafter PT23).

Since those papers, we have improved our likelihood function. While previously any evolution that ended with an eccentricity below the period-eccentricity envelope (Section \ref{subsec:bayesianlike}) was equally likely, we now take into account the probability density for final eccentricity; if a system ought to be circularized, sets of MCMC parameters that result in circularization are considered several orders of magnitude more likely than those that just fall under the envelope.

For this analysis, we are using the catalogue of 728 binary stars provided by \citet{2019MNRAS.489.1644W} (hereafter W19). The authors of that paper took light curve (LC) data from the Kepler mission \citep{kepler} and combined it with photometric information from various other sources. They then used Bayesian analysis to constrain their orbital and stellar parameters and identify eclipsing binaries (EBs) in the data.

The remainder of this article describes our analysis of those data. In Section \ref{sec:methods} we describe our methods, and in Section \ref{sec:Data} we explain what specific data were used. We present and analyze our results in Section \ref{sec:results}. Finally, our conclusions can be found in Section \ref{sec:conclusions}.

\section{Methods} \label{sec:methods}

\begin{figure}[tp]
\plotone{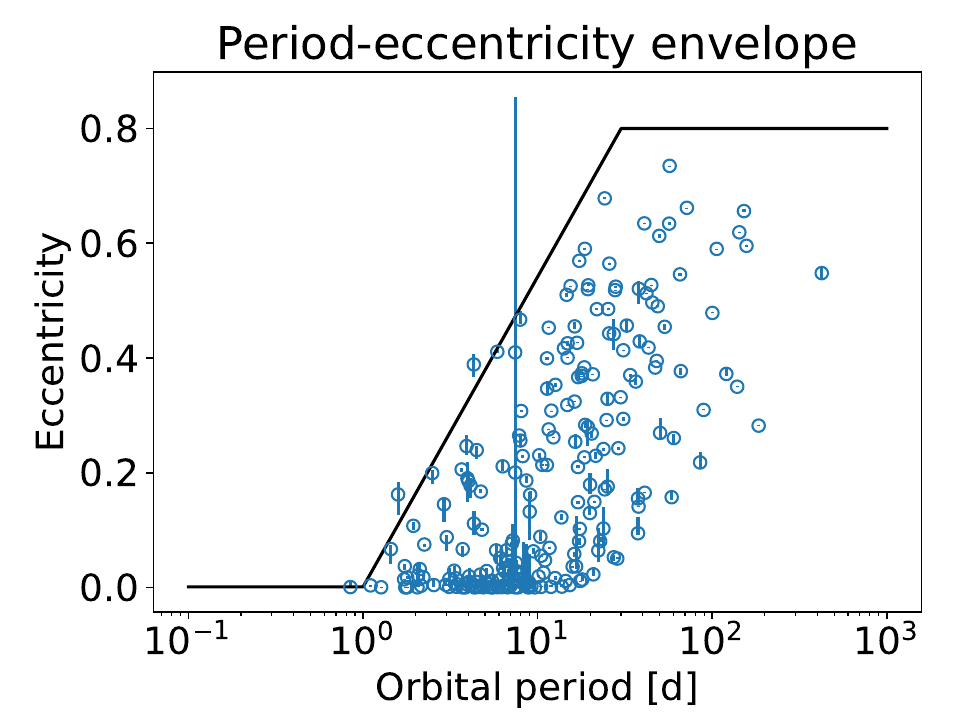}
\caption{Eccentricity vs. period for Kepler systems identified by
W19. The
period-eccentricity envelope indicates the eccentricity above which
no systems at a given period are observed. The long blue line is an error bar from a particularly poorly-constrained system.
\label{fig:peplot}}
\end{figure}

Our approach to constraining tidal dissipation relies on MCMC.
In each step, we calculate the orbital evolution for a set of tidal parameters, solving for some initial conditions as needed.
We compare the results to observed, present-day systems, and thus evaluate the likelihood of the input parameters.
The process is explained further in Sections \ref{subsec:bayesianlike}, \ref{subsec:POET}, \ref{subsec:tidemodel}, and \ref{subsec:emcee}. In Section \ref{subsec:prob_nn}, we describe the machine learning model we integrated into this process.

\subsection{Bayesian Likelihood} \label{subsec:bayesianlike}

Our likelihood function is
\begin{align*}
    \mathcal{L} = \frac{1}{\mathcal{Z}} \mathcal{D}_\theta \left( \vec{\theta} \right) & \mathcal{H} \left[ e_{env} - \hat{e}\left(e_i^{max},\vec{\theta},\vec{Q} \right) \right] \\
    & \times \Pi_\theta \left( \vec{\theta} \right) \Pi_Q \left( \vec{Q} \right) I, \numberthis
    \label{eq:lf1}
\end{align*}
\begin{equation}
    I \equiv \int_0^{e_i^{max}}\mathcal{D}_e\left[ \hat{e}\left(e_i,\vec{\theta},\vec{Q} \right) \mid \vec{\theta} \right] \Pi_e\left( e_i \right) de_i.
    \label{eq:lf2}
\end{equation}
We first explain Equation \ref{eq:lf1}. $\mathcal{Z}$ is the evidence, which we do not need to know.
$\mathcal{D}_\theta \left( \vec{\theta} \right)$ is the probability of observing what has in fact been observed if the correct physical parameters of the system, excluding eccentricity, are $\vec{\theta}$.
$\vec{Q}$ represents the dissipation parameters
(see Section \ref{subsec:tidemodel}). $\Pi_x \left( x \right)$ are the priors for the value $x$. $I$ is defined in Equation \ref{eq:lf2}.

We define a period-eccentricity envelope below which all systems fall (Figure \ref{fig:peplot}). We take this envelope to be indicative of dynamics these systems must be following.
We use $e_{env}$ to represent the maximum eccentricity a system of two stars with a given final period $P_{orb}$ could possibly have.
The value $\hat{e}\left(e_i,\vec{\theta},\vec{Q} \right)$ is the final eccentricity POET (Section \ref{subsec:POET}) predicts after an evolution with initial eccentricity $e_i$ which takes $\vec{\theta}$ and $\vec{Q}$ to be true, and assuming fixed final orbital period.
The Heaviside function $\mathcal{H}$ in Equation \ref{eq:lf1} enforces that $\hat{e}$ must fall below the envelope no matter the properties of the system.

We turn now to Equation \ref{eq:lf2}. Here, we integrate over all plausible initial eccentricities $e_i$, with $e_i^{max}$ the largest we want to consider. $\mathcal{D}_e\left( e \mid x \right)$ is the probability that $e_f=e$ if $x$ is true. $I$ is thus a weighted average of $\mathcal{D}_e$ over $e\in\left[ 0, \hat{e}\left(e_i,\vec{\theta},\vec{Q} \right) \right]$.

\begin{figure}[tp]
    \plotone{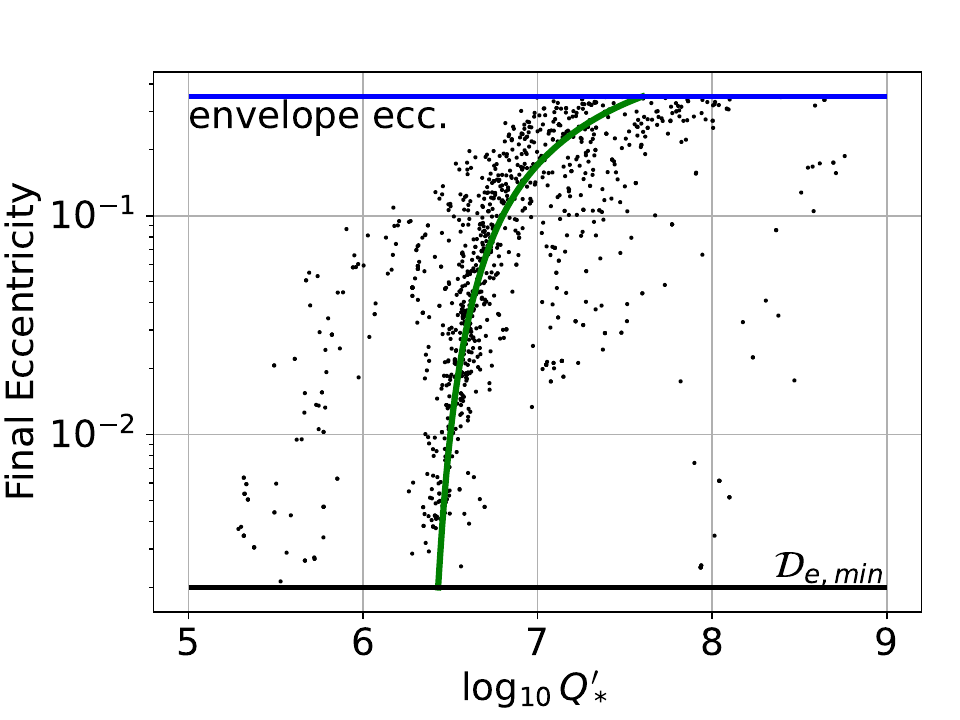}
    \caption{A demonstration of the boundaries we use for accepting or rejecting MCMC steps. Due to our high initial eccentricity, if dissipation is strong enough to drive final eccentricity below what is observed, the step must be rejected. If the dissipation is so weak that the system does not circularize below the envelope eccentricity, the step is rejected. Any combination of system and tidal properties (black points) that fall between these two boundaries would be advanced to the full likelihood calculation. The green line demonstrates a hypothetical result of holding all properties besides $\log Q$ constant.
\label{fig:circularization}}
\end{figure}

We take an initial eccentricity of $e_i^{max}=0.8$ to be the maximum
any of our systems might reasonably have had; higher values are unlikely to change our results and would be more computationally expensive. We begin every MCMC step (see Section \ref{subsec:emcee}) by finding the maximum final eccentricity of a system with properties $\vec{\theta}$, given $e_i^{max}$.
We note that if $e_f^{max}$ lands below the lower bound on observed eccentricity, the dissipation must be too high, as it should be possible to start with $e_i<e_i^{max}$ and evolve to the observed present day eccentricity.
Figure \ref{fig:circularization} demonstrates the range of valid parameters our process results in for a given system. The top line is the envelope eccentricity, and the bottom line represents the smallest final eccentricity supported by observation, $\mathcal{D}_{e,min}$. The points are individual sets of parameters accepted by our Bayesian analysis for further consideration. The green line, placed by hand, is a visual demonstration of the kind of trend we see when we hold all properties besides $\log Q$ constant.

We want Equation \ref{eq:lf2} in a form we can calculate.
Letting $e_i=\hat{e}^{-1}(e_f)$ and performing a change of variables from $e_i$ to $e_f$, we find
\begin{equation}
    I = \int_0^{e_f^{max}} \frac{\mathcal{D}_e(e_f\mid\vec{\theta})\Pi_e\left[ \hat{e}^{-1}(e_f) \right]}{\hat{e}'\left[ \hat{e}^{-1}(e_f) \right]} de_f.
    \label{eq:likint}
\end{equation}
We need to know $\hat{e}'$, the derivative of the final eccentricity with respect to initial eccentricity.
We make one of several approximations of $\hat{e}$ in order to find it; these are detailed in the Appendix.

This approach to the likelihood function takes advantage of tight constraints on observational data. Systems that should be circularized at longer periods will be reliably circularized in our evolutions.
The likelihood is proportional to the range of initial eccentricities which maps to the observationally determined range of final eccentricities.
For example, consider a system with observed $e_f\leq10^{-4}$.
Parameters which reproduce this level of circularization starting with $e_i = 0.1$ would be 100x more likely than parameters which require $e_i=0.001$.

\subsection{Orbital Evolution} \label{subsec:POET}

Orbital evolution, required to calculate $\hat{e}$, is handled by open-source software called POET \citep{POET}. POET has many useful features, including
\begin{enumerate}
    \item Various combinations of planets and stars,
    \item Splitting objects into an arbitrary number of convective and radiative zones,
    \item Tracking properties (e.g. size, tidal dissipation, obliquity, etc.) for each zone,
    \item Angular momentum loss due to wind, and
    \item Stellar evolution.
\end{enumerate}
We model stars as having two zones: a radiative interior and a convective exterior. The angular momentum of each zone is evolved separately, and can be transferred between the two via mass exchange. Differential rotation between them is taken to decay exponentially. Spins are assumed to be aligned with the orbit.
For stellar evolution, POET takes advantage of Modules for Experiments in Stellar Astrophysics (MESA), as well as MESA Isochrones \& Stellar Tracks (MIST). The exchange of angular momentum between radiative and convective zones follows the models presented in \citet{wind3} and \citet{Irwin_et_al_07},
as does the loss of surface zone angular momentum to stellar wind. The latter follows
\begin{equation}
    \frac{d\vec{L}_\mathrm{conv}}{dt}
    =
    -K_w\vec{\omega}_\mathrm{conv} \min(\omega_{\mathrm{conv}},
    \omega_\mathrm{sat})^2
			\sqrt{\frac{R_* M_\odot}{R_\odot M_*}},
\end{equation}
where $K_w$ is the efficiency of angular momentum removal, $\omega_{sat}$ is the frequency cutoff above which the mass loss rate due to stellar wind saturates, and $\vec{\omega}_\mathrm{conv}$ is the convective zone angular velocity vector.
POET adapts the tidal evolution model provided by \citet{laitome}, expanding it to allow for eccentric orbits and applying it to each zone separately. Briefly, per \citet{laitome} and other papers e.g. \citet{theBigZahn}, tidal potential is expanded as a series of temporal Fourier terms:
\begin{align}
    U_{tide}(\vec{r}, t)
    \equiv & {}
    \frac{GM'}{|\vec{r}_{M'}|}\left(
        \frac{\vec{r}\cdot\vec{r}_{M'}}{\left|\vec{r}_{M'}\right|^2}
        -
        \frac{\left|\vec{r_{M'}}\right|}{\left|\vec{r} -
        \vec{r}_{M'}\right|}
    \right)\nonumber \\
    = & {}
    \sum_{m=-2}^2 \sum_{m'=0}^{\infty} U_{m,m'} e^{-i\Omega_{m,m'} t},\label{eq:plzwork} \\
    \Omega_{m,m'} \equiv &  m\Omega_\star-m'\Omega_{orb},\nonumber
\end{align}
where $\Omega_\star$ is the spin angular velocity of the tidally
distorted zone and $\Omega_{orb}$ is the orbital
angular velocity.
$U_{m,m'}$ is further expanded with spatial spherical harmonics, resulting in
an associated expansion coefficient $p_{s,m'}$, which we have pre-calculated for $s=-2,0,2$ and $0\leq m'\leq400$.
As eccentricity increases, more $p_{s,m'}$ coefficients with larger values of $m'$ are required.
A more detailed description is provided in PS22.

\subsection{Tidal Dissipation Model} \label{subsec:tidemodel}
We assume dissipation only occurs in the convective zone of a star, and that it occurs primarily during the main sequence (MS).
We use a broken-powerlaw for the dependence of dissipation on period, evaluated independently for each term in the tidal potential expansion described in Section \ref{subsec:POET}.
We parameterize our tidal dissipation model as

\begin{equation}
    Q'_{m,m'} = Q_0 \max \left[ 1, \frac{P_{m,m'}}{P_0}^\alpha \right],
    \label{eq:model}
\end{equation}
where $Q_0$ is the maximum dissipation, $P_0$ is the period at which dissipation saturates, $\alpha$ is the powerlaw exponent, and $P_{m,m'}\equiv2\pi/\Omega_{m,m'}$ is the tidal period.
The first three are constrained using Bayesian analysis (Section \ref{subsec:bayesianlike}).
These parameters are taken as constant over the life of the system, which is reasonable due to our assumption that the MS dominates. The prescription is evaluated for each tidal term separately.
Because we do this for each system individually, Equation \ref{eq:model} need only fit the narrow range of $P_{m,m'}$ the current EB is most sensitive to (see Section \ref{subsec:bysystem}); it is thus flexible enough to capture virtually any smooth dependence of $Q'_{m,m'}$ on tidal period.
It is important to note that we are not sensitive to small periods $\ll$ 1 day. Also, since this prescription is used over the lifetime of a system for which we only know the final state, it is essentially an average. We capture inertial mode models only approximately, through their frequency-smoothed dissipation, and we do not include spin-dependence.

\subsection{Bayesian Implementation} \label{subsec:emcee}

To perform MCMC, we use emcee \citep{emcee}, a Python software package implementing the algorithms described in \citet{emmi}. We use 64 walkers whose starting positions at each step are derived from starting positions from the last step. In order to determine burn-in—when the quantiles for the tidal parameters have converged—we follow a version of the \citet{mcmcQuantiles} algorithm which has been adapted for emcee's multi-walker sampling.
More detail can be found in PS22 and \citet{itme3}.

During the course of an MCMC iteration, we need to know initial values for the system at two points.
First, we want $p_i$ when we impose the envelope on an evolution with $e_i^{max}$—the 1D case.
Though our desired output is $e_f^{max}$, it must correspond with a $p_f$ that agrees with observations.
Second, we want both $p_i$ and $e_i$—the 2D case—in two of our three approximations for $\hat{e}$ (Appendix \ref{apx:lik}). Here, the system should finish its evolution at a period and eccentricity that both fall within the constraints of observational data.
In either case, we cannot know in advance what the system actually started with, and we cannot simply run an evolution backwards because doing so is numerically unstable. Even in the case where we are only solving for the initial period, numerical instability prevents us from running backwards.
The only reliable way to find initial values is to solve for them numerically, which requires multiple additional evolutions to explore different initial conditions. We have two different solvers, 1D for the case where we only need the initial period, and 2D for when we need both the initial period and eccentricity such that the orbit evolves to the currently observed configuration.

\subsection{Machine Learning} \label{subsec:prob_nn}
\begin{figure}[tp]
    \centering
    \includegraphics[width=0.45\textwidth, height=0.7\textwidth]{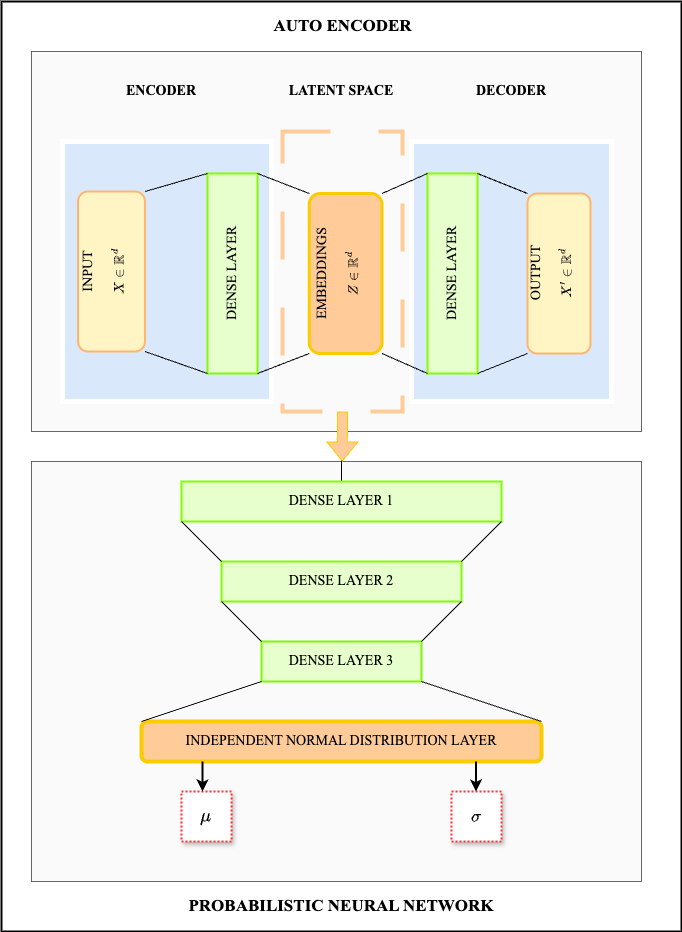}
    \caption{Overview of the deep probabilistic neural network architecture. The input data is processed by an autoencoder, $\mathcal{F(\cdot)}$, to learn a feature representation in an equi-dimensional latent space, preserving information (upper block). These features serve as input to the probabilistic neural network, $\mathcal{G(\cdot)}$, that generates the mean and standard deviation parameters for independent normal distribution layer (lower block).}
\label{ml_arch}
\end{figure}

As discussed in Section \ref{subsec:emcee}, we use two different numerical solvers to find initial values of period and eccentricity.
The 1D solver must have bounds, which in the past we have found by repeatedly trying evolutions, varying the initial period until the resulting final period crosses over the correct value.
The 2D solver relied on multiplying the final eccentricity by a constant to find a guess for a larger initial eccentricity.
Both cases benefit from improving the accuracy of the initial estimates, which can significantly reduce the number of iterations, and thus computational time, required for a solver to converge. We have developed a probabilistic neural network (PNN) to attempt to provide reasonable estimates for the initial values. The network architecture comprises an autoencoder followed by a deep PNN as shown in Figure \ref{ml_arch}.

The autoencoder is a single-layer encoder-decoder model. The input data are initially preprocessed through a custom scaling procedure comprising mean centering and range scaling. Subsequently, these scaled data, $x \in X$, where $X \in \mathbb{R}^{d}$, serve as an input to the encoder layer, which effectively maps them to equidimensional latent space embeddings, $z \in Z$, where $Z \in \mathbb{R}^{d}$. These embeddings are transmitted to the decoder layer to reconstruct the original input, $x' \in X'$, where $X' \in \mathbb{R}^{d}$, by optimizing the trainable parameters using mean squared error as the loss function. Even with its relatively straightforward architecture, the autoencoder effectively carries out nonlinear feature extraction, transforming the original data into new, more informative features within the latent space. The autoencoder is designed to encapsulate the critical elements of the data in a more compact and meaningful manner, thereby eliminating degenerate cases that may lead to numerical instability. Furthermore, it implicitly scales the data into a more appropriate distribution for the probabilistic model. The embeddings from the latent space are then input into the PNN, which comprises three dense layers followed by an independent normal distribution layer. This network learns the statistical parameters, i.e., $mean (\mu)$ and $stdev (\sigma)$, using a negative log-likelihood loss function to estimate suitable initial guesses for the solver. The mathematical representation of the neural network model is as follows:
\begin{equation}
 \begin{split}
    z = \mathcal{F}(x), \quad x\in X \\
    \mu, \sigma = \mathcal{G}(z), \quad z \in Z
 \end{split}
\end{equation}
where $\mathcal{F}(\cdot)$ is the autoencoder, $\mathcal{G}(\cdot)$ is the PNN model, and $(X, Z) \in \mathbb{R}^d$.

We build the training dataset from the stellar and system parameters, and the resulting final values, of every evolution attempted during the MCMC process. The data are partitioned by individual system; this prevents the PNN from collapsing due to the excessive variance in the parameter space that may arise when combining heterogeneous systems. For each system, we train three distinct models: 1D period, 2D period, and 2D eccentricity. The first is used when only the initial period is needed. The latter two are for the 2D solver; they're trained on the same input features, but with either period or eccentricity as the output. Finally, all three of a system's datasets are further subdivided at the median of the target output, to simplify the behavior any single PNN needs to predict; this gives us a total of six models per system. At the start of each two-day computational run, we train our models on all data accumulated to that point.

The PNN output is given in terms of upper and lower bounds $2\sigma$ above and below the mean. We constrain the lower bound to be strictly $>0$, and, for eccentricity, the upper bound to be $\leq0.8$. For the 2D solver, we use the mean of these bounds as the initial guess. For the 1D solver, the bounds are used as lower and upper limits. If the solver fails because the bounds predicted by ML do not in fact enclose the solution, we automatically fall back to our old method of deriving the bounds, running yet more evolutions before we begin the actual solver. That approach is guaranteed to bound the correct initial value, though at a potentially significant computational cost.

\section{Data} \label{sec:Data}

All of the data for the binaries analyzed in this work come from W19.
Per that paper, the results of some systems were questionable, namely those with anomalously young age or
large enough morphology, $morph = (R_1 + R_2) / a$, 
as these are likely to have undergone mass transfer, breaking their assumptions. Other systems break our assumptions; for example,
non-Sunlike stars (for which dissipation may behave differently).
We thus removed multiple EBs from that initial dataset.

Specifically, we avoid systems with surface gravity less than $10^4$ $\mathrm{cm}/\mathrm{s}^2$,
age less than $316$ $\mathrm{Myr}$,
and morphology $\geq0.5$.
Also, we are limited to handling systems that fit within the evolution tracks we pre-computed (see Section \ref{subsec:POET}). For this reason, we exclude all systems with solar masses outside the range $\left[0.4, 1.2 \right]M_\odot$ and
metallicity outside $\left[-1.014, 0.537 \right]$.
This left us with 127 systems.

Two systems (KIC 11234677 and KIC 6301030) failed to find starting positions. Others did not converge in time, because they either needed too many steps or took too long to take steps, or both. Four systems were delayed because their current step was taking longer than our maximum job time, necessitating a reseed so that MCMC would choose different system parameters. In total, there were 22 systems that did not finish, leaving 105 that we analyze.

\section{Results} \label{sec:results}

\subsection{Machine Learning Performance}  \label{subsec:mlperf}
We tested the performance of our PNN using 70 of the systems which had not yet converged at the time of testing. For each of these systems, we randomly selected 120 sets of values used to prompt the period and eccentricity solver which the associated models had \textit{not} trained on. We split those values into eight parallel-processed groups of 15. We then timed how long it took the solver to finish computing these tests both with and without our probabilistic model.

\begin{figure}[tp]
    \centering
    \includegraphics[width=0.45\textwidth]{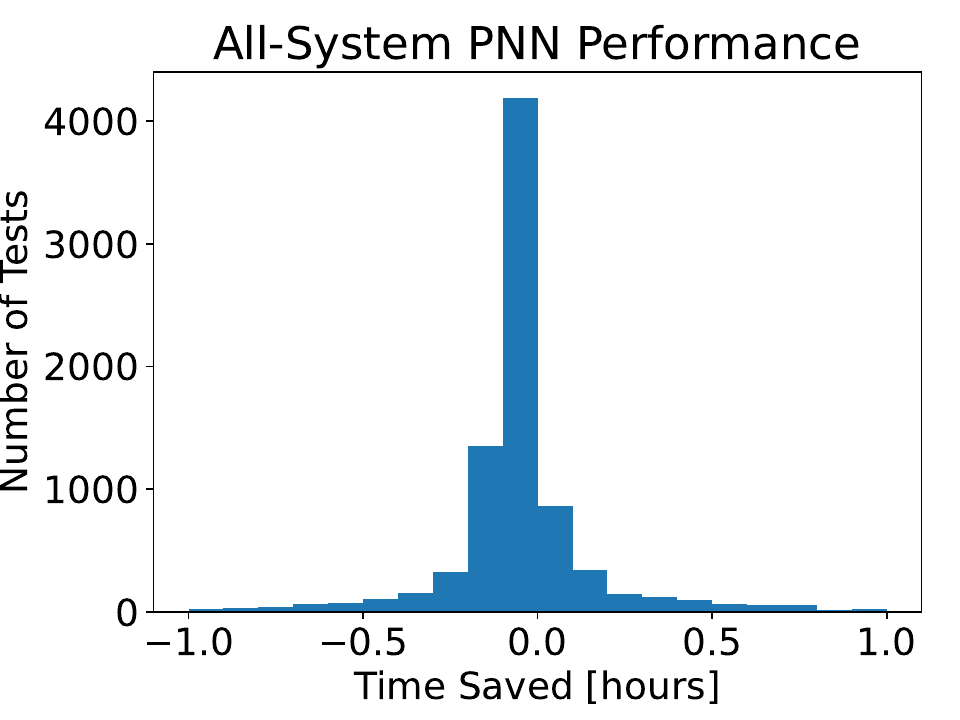}
    \caption{Histogram of performance results from almost every test performed. Time saved is time to solve without PNN minus time to solve with. Some parts of the range are not displayed in order to show more detail.}
\label{fig:allml}
\end{figure}

\begin{figure}[tp]
    \centering
    \includegraphics[width=0.45\textwidth]{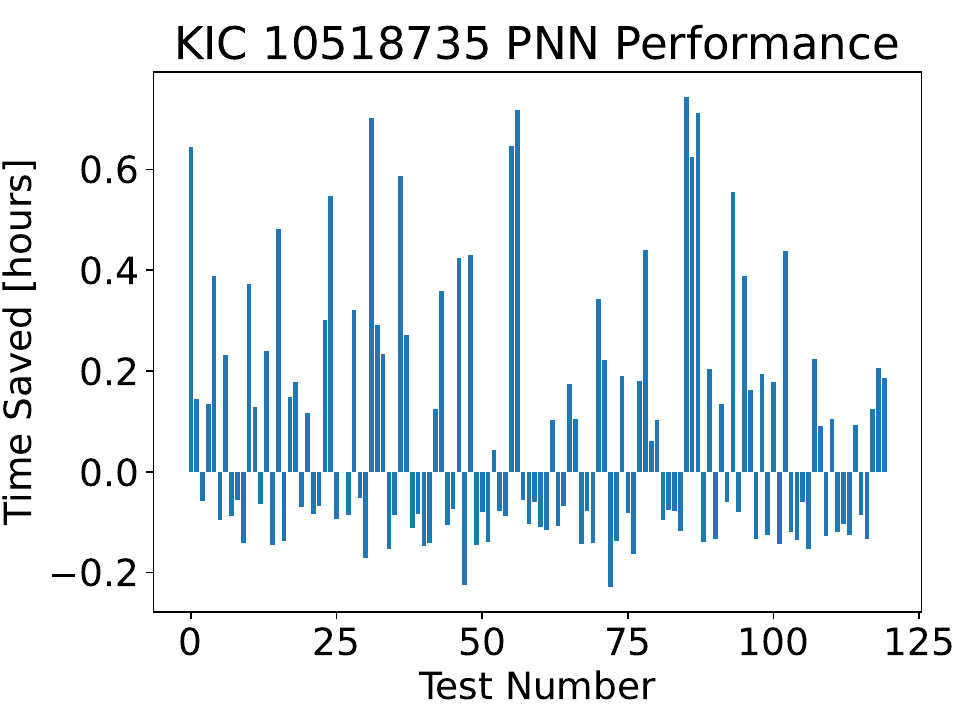}
    \caption{Example result from one system. Each test solve was run twice, once with PNN and once without. Time saved is calculated as in Figure \ref{fig:allml}.}
\label{fig:specificml}
\end{figure}

Our machine learning implementation had an uneven, somewhat deleterious effect on solving times. On average, an individual test took just over two minutes longer when using the PNN for predictions versus not. Only 18 of the 70 systems showed a time savings. However, this does also speak to the ambiguity of the performance impact; a higher percentage of systems with an average time savings saw a difference of $\geq$ five minutes compared to those with a loss (10 out of 18 vs. 22 out of 52). The best-performing test average saved just under 30 minutes, while the worst-performing lost just under 20 minutes.
Figure \ref{fig:allml} is a histogram of almost every test solve performed (the x-axis is zoomed in to show more detail); it clearly shows a bias towards time losses, but a significant number of instances where time is saved as well. On the individual system level, 45 systems are either all or mostly time losses, or see great enough time losses to negate the tests which saved time. Ten systems are likewise preferentially saving time. The results of the remaining 15 vary widely between tests (e.g. Figure \ref{fig:specificml}), with practical time savings or loss determined primarily by what values MCMC randomly selects.

\subsection{Machine Learning Analysis} \label{subsec:mla}

There does not appear to be a relation between model performance and system properties. It is possible that there was a selection effect stemming from using systems that have taken longer to converge for this test. However, longer convergence times are the result of necessities in numerical calculation, rather than differing physics that would degrade the PNN's performance. In addition, because the final test systems comprise more than half of our total systems, any impact is likely minimal.
Another possible issue is the composition of the training data, which includes every evolution performed. Because the evolutions requested by the solvers incrementally change one (two) variable(s) out of 10 (11), the dataset becomes dominated by many extremely self-similar islands in parameter space, with potentially large gaps between them. This could prevent the PNN from generalizing the underlying physics, hindering its ability to interpolate between these regions.
It might also be the case that certain hyperparameter choices may have restricted performance. Specifically, splitting the data on the median may have been suboptimal, and perhaps using a different quantile, or even choosing where we partition on a system-by-system basis, would have improved the results.
Finally, it's possible that the chosen neural network architecture was simply unable to capture the physics behind this particular problem.

\subsection{Individual Constraints} \label{subsec:bysystem}

\begin{figure}[tp]
    \centering
    \includegraphics[width=0.45\textwidth]{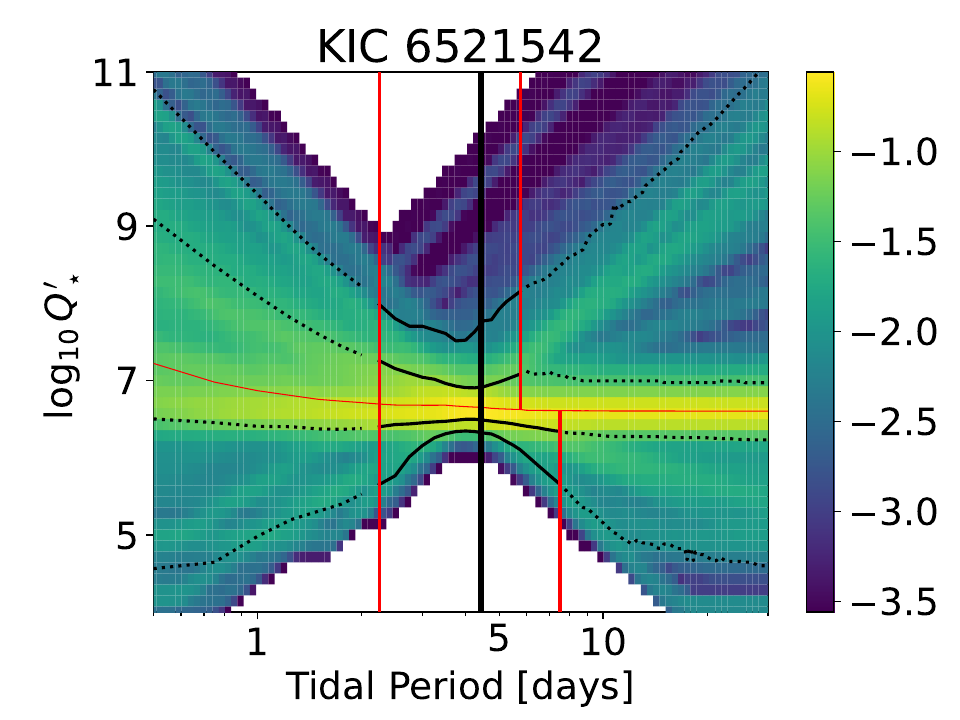}
    \caption{Tidal constraints for KIC 6521542. Lighter colors indicate higher probability; the color bar follows a log scale. The black vertical line is present-day orbital period. The red vertical lines bound the part of the data which is driven by system dynamics. The four black curves indicate the 2.3\%, 15.9\%, 84.1\%, and 97.7\% quantiles; the red curve shows the median.}
\label{fig:individ_const}
\end{figure}

We have found individual constraints for 105 systems. An example is shown in Figure \ref{fig:individ_const}.
In that graph, lighter colors indicate higher probability of that combination of values. The black vertical line is the system's present-day orbital period. The four black curves indicate the 2.3\%, 15.9\%, 84.1\%, and 97.7\% quantiles, while the red curve shows the median.

The red vertical lines bound the part of the constraint which is driven by system dynamics; outside of them, the slope of the upper (lower) quantile matches the limits we place on the prior for $\alpha$. This indicates that the results become dominated by the prior rather than the data.
We take the points this transition occurs as being when the top (bottom) quantile becomes greater (less) than its minimum (maximum) by more than 0.7 dex.
When determining the burn-in period, we only need it to be long enough for the quantiles \textit{within} these ranges to have converged, since we ignore the distribution outside of them.

The upright and inverted ``v" shapes in the graph indicate the tightest constraints on tidal dissipation.
These are primarily a product of two requirements: first, that we fall underneath the period-eccentricity envelope, and second, that we not end up more circularized than observations.
These are not necessarily centered on the same tidal period, as they come into play at different times during the system's evolution.

It is not always possible to find both top and bottom constraints for a system. Once dissipation has met the critical value needed to circularize a system, eccentricity can not be decreased any further, and so there is no observable difference with this method; circularized systems thus only have an upper constraint.

\subsection{Combined Constraints} \label{subsec:combined}

\begin{figure}[tp]
    \centering
    \includegraphics[width=0.45\textwidth]{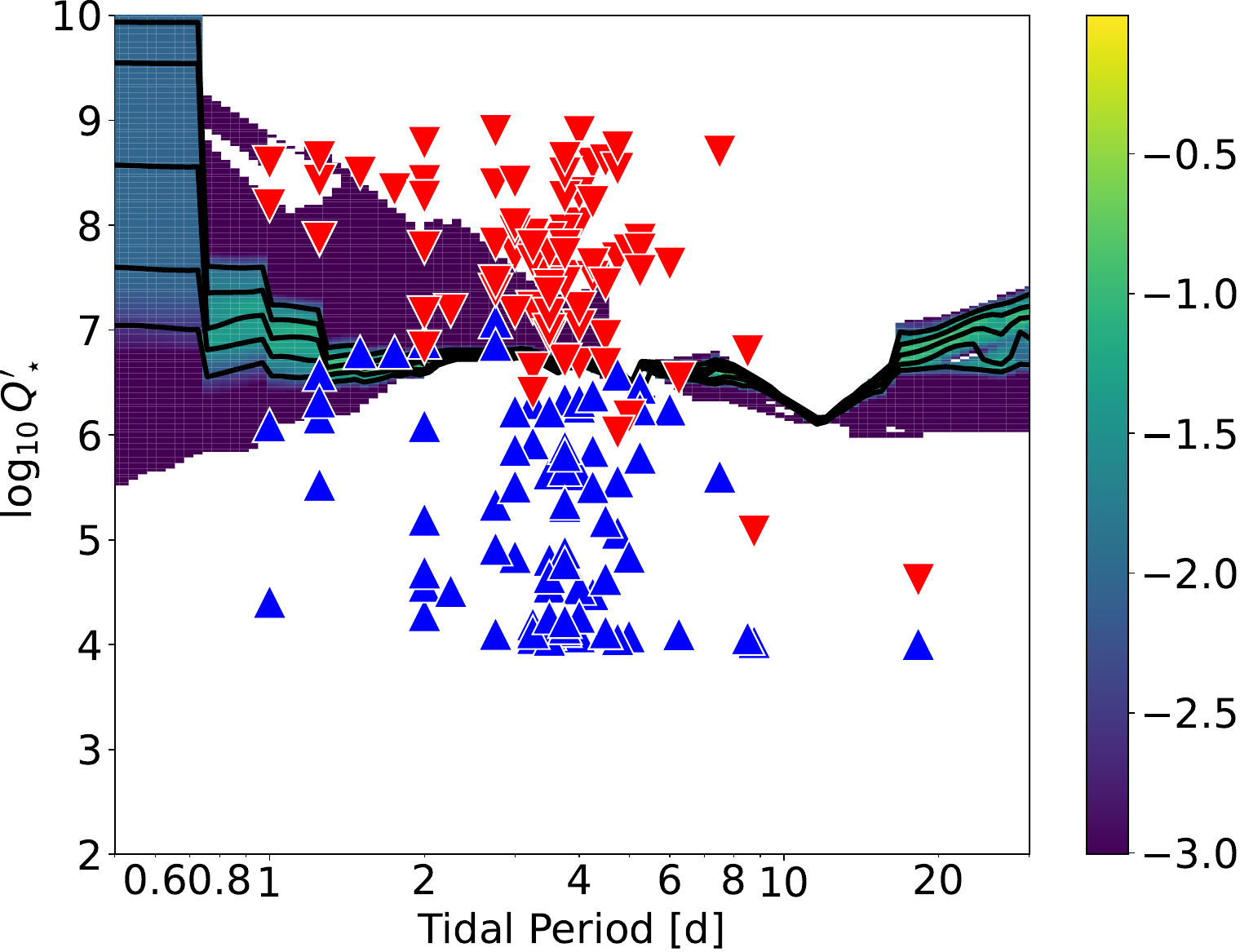}
    \caption{Combined constraints for the majority of systems. Triangles indicate tightest bounds of individual systems: blue triangles mark the lower quantile, while red inverted triangles show the upper. All other graph elements are as in Figure \ref{fig:individ_const}. 24 systems were skipped.}
\label{fig:comb_const}
\end{figure}

By combining individual constraint posteriors, we can find a group constraint that agrees with all individual systems.
To accomplish this, we find, for each tidal period, the systems with red lines that bound that period, and multiply those systems together.
Figure \ref{fig:comb_const} shows the resulting combined constraint of $\log Q'_* \approx 6.75$; it is formatted the same as Figure \ref{fig:individ_const}, with the addition of arrows marking the tightest constraints of the included individual systems. These arrows, which indicate the
2.3\% and 97.7\%
quantiles at the point of a system's tightest constraint,
demonstrate that the large majority of systems included in this graph are in clear agreement with the combined constraint.
Most of the handful of outliers are close enough to still have overlap between their individual constraints and the combined constraint.
Two systems (KIC 11232745 and KIC 11704044) stand out (the red and blue triangle pairs in the lower right). These were not automatically rejected because, though their tightest constraints do not agree with the group consensus, some area within their posteriors does.
These two systems are very circularized: maximum likelihood values from \citet{2019MNRAS.489.1644W} are $e=5.3\times10^{-8}$ and $2.9\times10^{-6}$, respectively, with median and one-sigma errors $2.0\times 10{^{-4}}^{+1.5\times 10^{-3}}_{-1.7\times 10^{-4}}$ and $1.45\times 10{^{-3}}^{+2.7\times 10^{-3}}_{-1.43\times 10^{-3}}$. In such a case
our likelihood function strongly prefers values of the tidal dissipation that can drive a wide range of initial eccentricities down.

Figure \ref{fig:comb_const} has only 81 systems. Two groups of systems were removed in order to create it.
First, ten systems in our set of 105 (KICs 4753988, 4948863, 5039441, 6525196, 6546508, 9715925, 10268903, 10483644, 9110346, and 5731312)
were identified in \citet{gawdsend1} as having a third star associated with them. Though these were on quite distant orbits, we can still expect to see von Zeipel-Lidov-Kozai oscillations \citep{gawdsend2}, and the timescale for these is short enough to break our assumptions. Thus, these systems have been excluded.
The other fourteen systems (KICs 9775253, 10385682, 10091257, 11252617, 3973504, 7369523, 11616200, 11200773, 4276114, 10935310, 5288543, 11071207, 6697716, and 7798259) were rejected automatically by our combining process, as they did not overlap with the combined constraint formed from the majority of the systems. \citet{gawdsend1} found almost $10\%$ of Kepler binaries to clearly have associated third bodies, with significant likelihood that a higher fraction is a more accurate picture; this can reasonably explain the $\approx23\%$ of our systems which seem to not follow the general trend.

\subsection{Analysis} \label{subsec:analysis}

We now examine how our results compare with notable tidal dissipation models.
Equilibrium tide models, such as \citet{old2} and \citet{starStar3}, expect that $Q \propto P_{tide}^{-1}$. As is shown by Figure \ref{fig:comb_const},
the combined constraints we have found in this paper remain flat rather than demonstrating such a relation; however, it is not unexpected that we would see disagreement here, as the mechanism these papers address is weak in the fast tides regime and should not be able to provide the circularization we see.
Similarly, models that incorporate breaking g-modes, such as \citet{2010MNRAS.404.1849B}, say that dissipation should go as the 8/3 power of tidal period. This is also clearly not the behavior we see.
Inertial mode models, such as \citet{premsvsms}, \citet{2022ApJ...927L..36B}, and \citet{2026ApJ...997...27D}, generally indicate a period-dependent tidal dissipation, but, as stated in Section \ref{subsec:tidemodel}, we are only sensitive to these models' frequency-smoothed dissipation; only through combining multiple systems in the group constraint do we have a possibility of detecting dependence.

\begin{figure}[tp]
    \centering
    \includegraphics[width=0.45\textwidth]{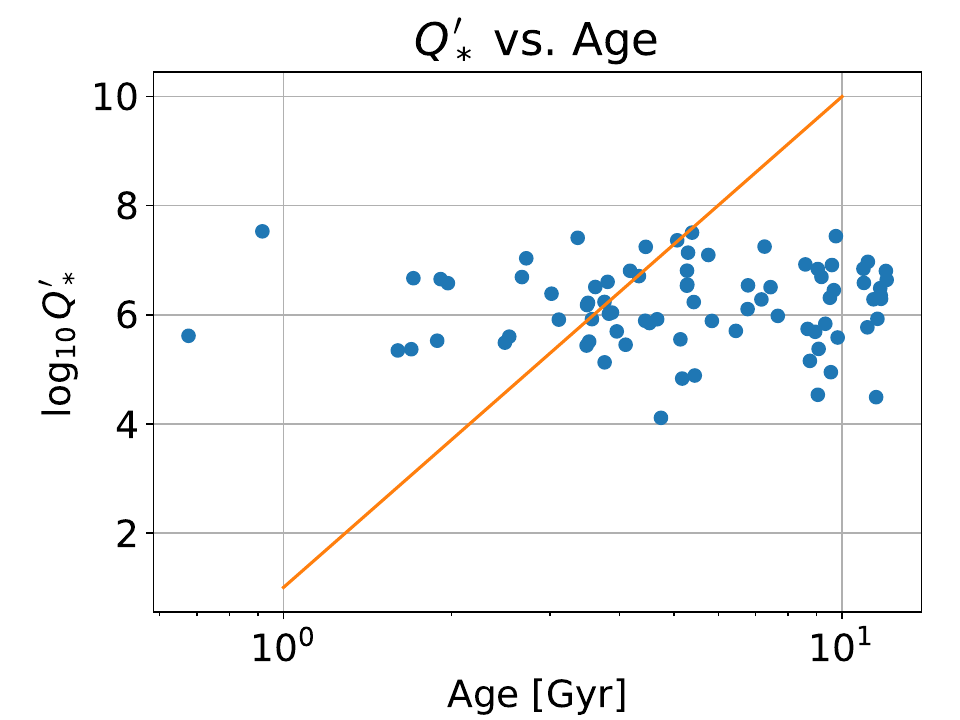}
    \caption{Median $Q_*'$ versus age. Blue points are systems included in the group constraint. The straight line indicates what we would expect to see if circularization on the pre-MS dominates over MS dissipation, per \citet{premsvsms}.}
\label{fig:prems}
\end{figure}

A common prediction in many models is that
stars will have different dissipation between their pre-MS and MS phases, varying with age, mass, and spin; the values we have measured, then, would actually be
averages over the life of the system.
If pre-MS dissipation dominates, we would expect to see $Q$ increase with age, showing a decrease in measured (average) dissipation; this is because our analysis assumes MS dissipation dominates and hence older systems will have more time to circularize requiring less dissipation (higher $Q$) to achieve the orbit that in reality would have been set during pre-MS. However, when we graph each system's median $Q$ at the best-constrained period versus age, we see no evidence of a relation (Figure \ref{fig:prems}).

\begin{figure}[tp]
    \centering
    \includegraphics[width=0.45\textwidth]{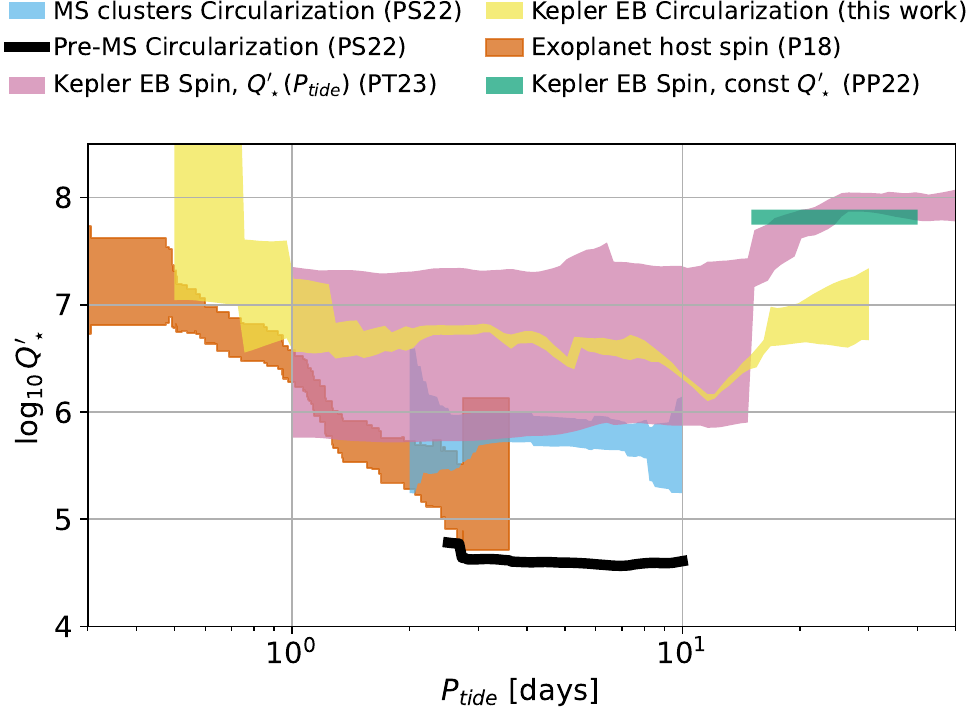}
    \caption{Current and previous results produced by our group; PP22 is \citet{2022MNRAS.512.3651P}. Each colored area is a combined constraint.}
\label{fig:previouslyon}
\end{figure}

The combined constraints we have found appear to be in tension with past results by our group, despite large similarity in our assumptions and procedures. As Figure \ref{fig:previouslyon} shows, for tidal period between 1 and 10 days the main area of overlap with previous constraints is with PT23, but in this range the priors dominated for that work. There is \textit{some} notable agreement with \citet{2018AJ....155..165P} (P18) for the lowest periods, but this is lost past $P_{tide}=1$ day. Also, PS22 found dissipation for a pre-MS cluster to be significant.

\begin{figure}[tp]
    \centering
    \includegraphics[width=0.45\textwidth]{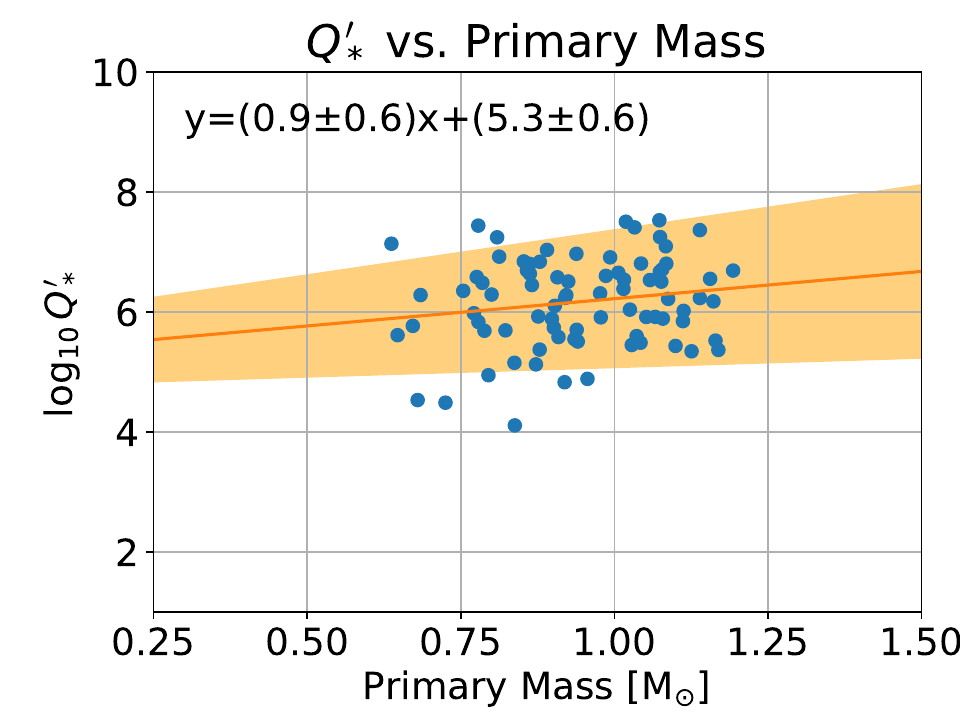}
    \caption{Median $Q_*'$ versus mass. Blue points are systems included in the group constraint. The equation describes the best-fit trend line, graphed in orange; the shaded orange area shows the bounds of the one-standard-deviation errors.}
\label{fig:qvmass}
\end{figure}

We have investigated and ruled out three possible causes for this discrepancy.
First, in Figure \ref{fig:qvmass}, we check to see if there is a relation between $Q$ and the mass of the primary star. If there were a relation, it would be possible for a difference in the mass distribution of two sets of systems to result in different group constraints. We find a best-fit trend line; the standard deviation for the slope is more than half the slope itself. There is no statistically significant relation between these values.
Second, our previous papers (PS22, etc.) used spectroscopic binaries and thus had much lower precision on their eccentricity measurements. Nonzero eccentricities $\lesssim0.05$ were indistinguishable from circular. Such systems do not provide a lower limit on $Q$. However, in the W19 dataset such low eccentricities are distinguishable, providing lower limits which could potentially push our constraints to higher values. We tested this by discarding the lower bounds on $\log Q_*$ for systems with observed $e<0.05$ to see if this would alleviate the issue. Though the lower bounds did fall (to $\log Q \approx 6$ at $P_{tide} < 1$ and $\log Q \approx 6.5$ at $P_{tide} < 12$), there was not a significant increase in overlap.
Third, we note that PS22 worked with binary stars in open clusters. Fewer systems were available per cluster than we had access to using the Kepler binaries (40 for NGC 188 versus 127), and, quite notably, the period-eccentricity envelope for the W19 systems was significantly higher than those for the open clusters. We considered the possibility that small-number statistics led to the use of inaccurate PE envelopes for those groups. We randomly sampled 40 W19 systems (the same number as that used to build the NGC 188 envelope) and determined how many were above the NGC 188 envelope and how many were below. Out of 10,000 iterations of this test, only 39 draws had no systems above the envelope. It is thus unlikely for this to explain the discrepancy.

\begin{figure}[tp]
    \centering
    \includegraphics[width=0.45\textwidth]{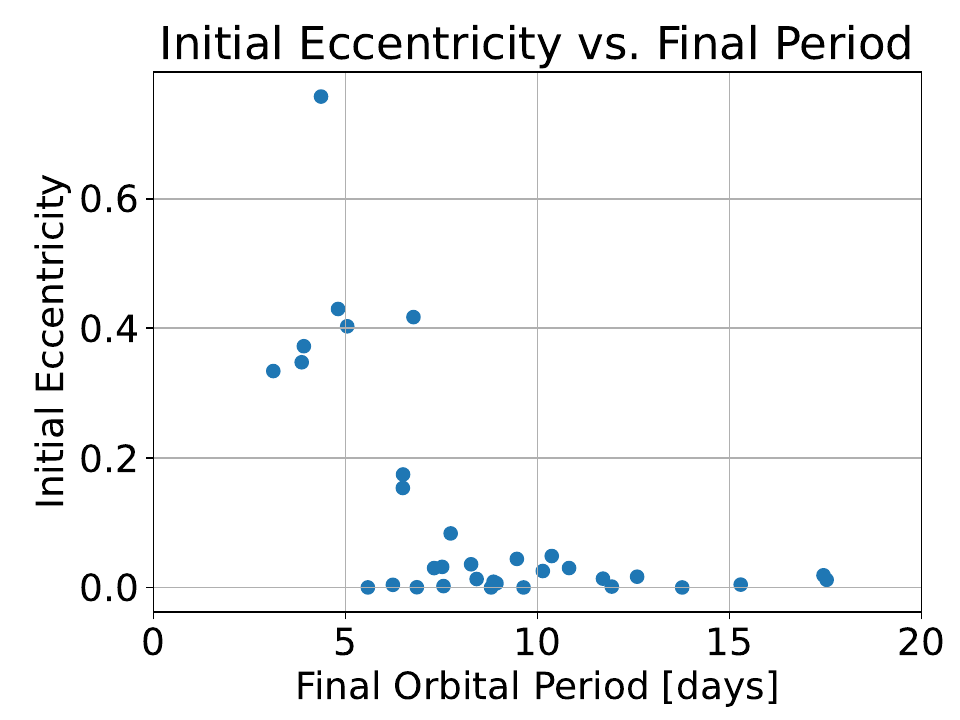}
    \caption{Initial eccentricity vs. final orbital period for present-day circularized systems, as calculated by 2D solver using maxlike parameters and flat stellar dissipation with $Q'_*=6.5$. Cold core is seen for period $\lesssim6$ days.}
\label{fig:zanz}
\end{figure}

It is possible there is a difference in behavior between binaries associated with open clusters and the overall set observed by Kepler. Open clusters start off as much denser environments than most W19 stars were likely born in. As a result, it is possible that the latter have more third bodies than the former (Section \ref{subsec:caveats}), as those in the open clusters would more likely have been ejected due to flybys.
If this is the case, it could potentially alter the dynamics of our group constraint, with for example closer-in binaries being less directly impacted by the third body than those further away.

We now turn to a finding by \citet{starStar1}, who show that in the range $3\lesssim P_{orb} \lesssim 6$ there exists two populations: a cold core of systems that circularize quickly, and an envelope consisting of systems that take longer.
In Figure \ref{fig:zanz}, we plot, for 33 systems with $e_f<0.05$, their initial eccentricity (as determined by our 2D solver when given maxlike values for orbital and stellar parameters and a flat $Q=6.5$) versus their observed period. We see, for low enough period, large $e_i$; we are thus capable of explaining cold core up to a period of around six days, while at the same time reproducing the much larger eccentricities of the systems in the envelope. This is possible due to our improved likelihood taking the initial eccentricity into consideration.

\begin{figure}[tp]
    \centering
    \includegraphics[width=0.45\textwidth]{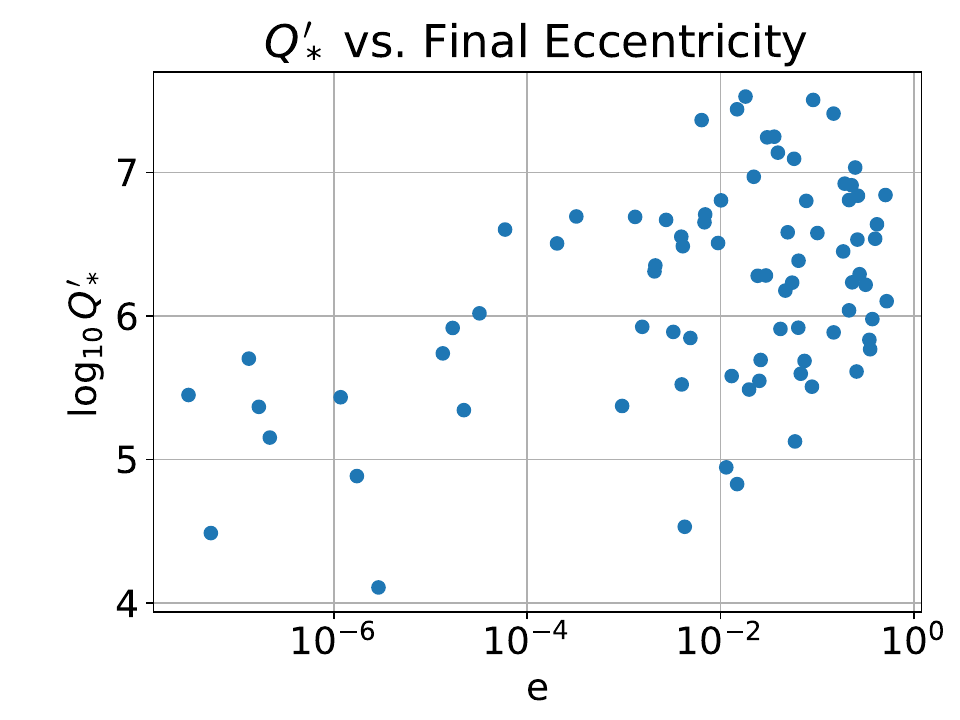}
    \caption{Median $Q_*'$ versus final eccentricity. Blue points are systems included in the group constraint.}
\label{fig:heck}
\end{figure}

Finally, we examine the possibility that $Q$ has dependence on spatial frequency waves $m$, $m'$ (Equations \ref{eq:plzwork} - \ref{eq:model}). In Figure \ref{fig:heck}, we graph median $Q$ versus final eccentricity. We see a small but clearly-present positive correlation; this is suggestive of higher spatial frequency waves having lower dissipation.

\subsection{Caveats} \label{subsec:caveats}

As already discussed, we likely have not excluded every system which is actually a triple. Undetected triples, while impacting the magnitude of dissipation, will not do so to an extent noticeable on a log scale if they are far enough away.
However, if they are close enough, the constraint we find for $Q$ will be significantly larger than it actually is. This has the potential to mask the true dynamics of the combined constraints.

Our results rely heavily on the presumed accuracy of our stellar evolution models; any inaccuracies on their part would impact the dissipation we find for a system.

As noted in Section \ref{sec:Data},
we threw out 22 systems that did not converge in the time available. It is possible that, rather than simply numerical issues, the increased difficulty in calculating these systems is a symptom of actual physics which might also have an impact on tidal dissipation, and that excluding them distorts our view of those dynamics. However, we checked the stellar and orbital properties of these systems, and they do not appear to be notably different from the rest of the sample.

\section{Conclusions} \label{sec:conclusions}

We have used Bayesian analysis to constrain the tidal dissipation in Kepler binaries.

We attempted, unsuccessfully, to use machine learning to speed up the constraint process. Considering the success other groups have had using machine learning in astrophysics (e.g. \citet{bandwagon1}, \citet{bandwagon2}, and \citet{bandwagon3}, to cite just a few examples), this is still a viable avenue for increasing efficiency.
Future efforts to improve our implementation should start with ensuring variety in the training data before moving on to changing the model itself.

We now compare to various predictions.
Using an equilibrium tide model, \citet{starStar3} finds  $9 \leq \log Q'_* \leq 10$ for solar-mass MS stars in the range of tidal periods we probe; they note that this is too high to explain the observed circularization, which we confirm.
Inertial wave models often have highly variable dissipation, so for the purposes of modeling circularization they often use a frequency-smoothed dissipation model. This results in $Q'_*$ independent of tidal period but scaling quadratically with spin period.
\citet{Barker_20} and \citet{2022ApJ...927L..36B} are examples of this approach; they predict $\log Q'_* \approx 7$, with some mass and age dependence. This is quite close to our group constraint of $\log Q'_* \approx 6.75$. In those papers, $Q'_*$ is roughly flat for near solar-mass stars of ages between 1 and 10 $\mathrm{Gyrs}$, which is the range most of our systems fall into. However, we do not agree well with their pre-MS dissipation, which as with many inertial models is much larger.
Finally, we see mixed results with resonance locking g-mode models. \citet{itsnotjustus} finds $\log Q'_* \approx 6.3$, though that paper also claims negligible circularization during MS. \citet{doorFour} aligns well with our results for $0.5\leq P_{tide} < 1$, falling as roughly $Q \propto P_{tide}^{-3}$ from $\log Q'_* \approx 8$ to $\log Q'_* \approx 7$, but above this their period dependence drives their predicted $Q'_*$ below our flat constraint.

We are capable of finding both the cold core and envelope described in \citet{starStar1} simultaneously.
We also do not see evidence of a division between pre-MS and MS dissipation. However, our results point towards the possibility that there are as-yet-undetected third bodies in many of the Kepler stars. If this is the case, the true dynamics may be hidden. Future observations may shed more light on this matter.

We are currently working on constraining stellar and orbital parameters for TESS systems (Schussler et. al., submitted). Finding tidal dissipation constraints for these systems is planned to occur in the future. When this is complete, they can be compared against the constraints we have found in this paper, potentially helping to clarify the origin of the discrepancy in $Q$.

\begin{acknowledgments}
This research was funded by NASA grant 80NSSC23K1486.
The authors acknowledge the Texas Advanced Computing Center (TACC) at The
University of Texas at Austin\footnote{URL: http://www.tacc.utexas.edu}
and High Performance Computing at The University of Texas at Dallas (HPC@UTD)
for providing HPC resources and support that have contributed to the
research results reported within this paper.
We are grateful to Haoyu Shen for identifying a crucial paper;
Megan Tran, Ashkan Jafarzadeh, Jessica Bell, and Brynne Menkhaus for proofreading;
and the reviewer for the insightful comments.
\end{acknowledgments}

\section*{Data Availability} \label{sec:datavab}

The version of POET used in this project can be found at \dataset[10.5281/zenodo.7742991]{https://doi.org/10.5281/zenodo.7742991}.

All outputs from this project are available at \dataset[10.5281/zenodo.21923497]{https://doi.org/10.5281/zenodo.21923497}. Contained within are the MCMC samples for each system, all quantiles for $\log Q$, the burn-in for those quantiles, and the related standard deviations.
The final forms of other programs and scripts used to create these outputs can be found at \dataset[10.5281/zenodo.21936961]{https://doi.org/10.5281/zenodo.21936961} and \dataset[10.5281/zenodo.21937069]{https://doi.org/10.5281/zenodo.21937069}.

\vspace{5mm}

\software{
          POET \citep{POET}, 
          MESA \citep{MESA,MESAb,MESAc,MESAd}, 
          MIST \citep{mist}, 
          emcee \citep{emcee},
          TensorFlow \citep{tensorflow2015-whitepaper}
          }

\appendix

\section{Likelihood Approximations} \label{apx:lik}

\begin{figure}[tp]
    \centering
    \includegraphics[width=\textwidth]{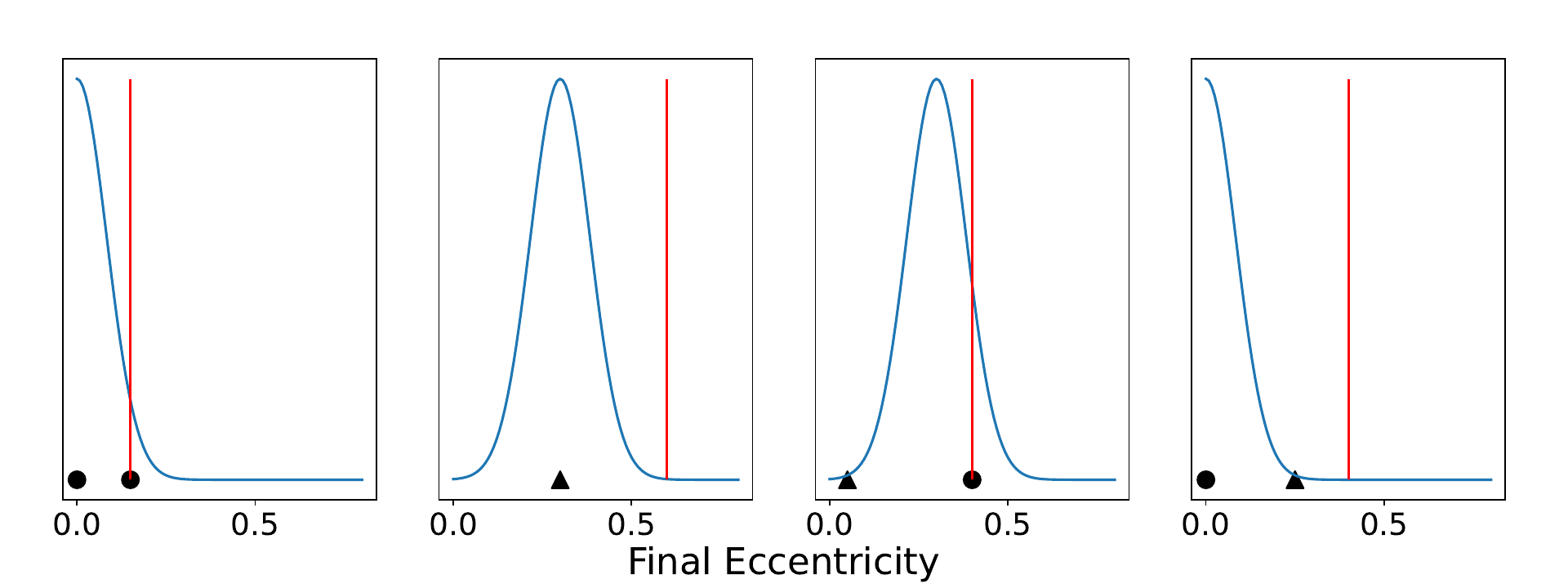}
    \caption{Different cases for our approximation of $\hat{e}(e_i)$. The blue curve represents $\mathcal{D}_e$; its width is exaggerated for visibility. The red vertical line is $e_f^{max}$. A circle indicates a point for which we already know both x and y values. A triangle indicates the point for which we calculate $e_i$ and slope in a given case.}
\label{fig:appx1}
\end{figure}

Here we describe our procedure for approximating the function $\hat{e}(e_i)$, which maps an initial eccentricity to the final eccentricity a system would arrive at by the present-day age given system parameters $\vec{\theta}$ and dissipation parameters $\vec{Q}$, and holding final period $P_{orb}$ fixed to what has been observed. The approximation need only be valid in the range of the observational distribution of present day eccentricities $\mathcal{D}_e$, where $\mathcal{D}_{e,min}$ is the lowest value of $e$ the distribution supports and $\mathcal{D}_{e,max}$ is the highest.

\begin{figure}[tp]
    \centering
    \includegraphics[width=0.45\textwidth]{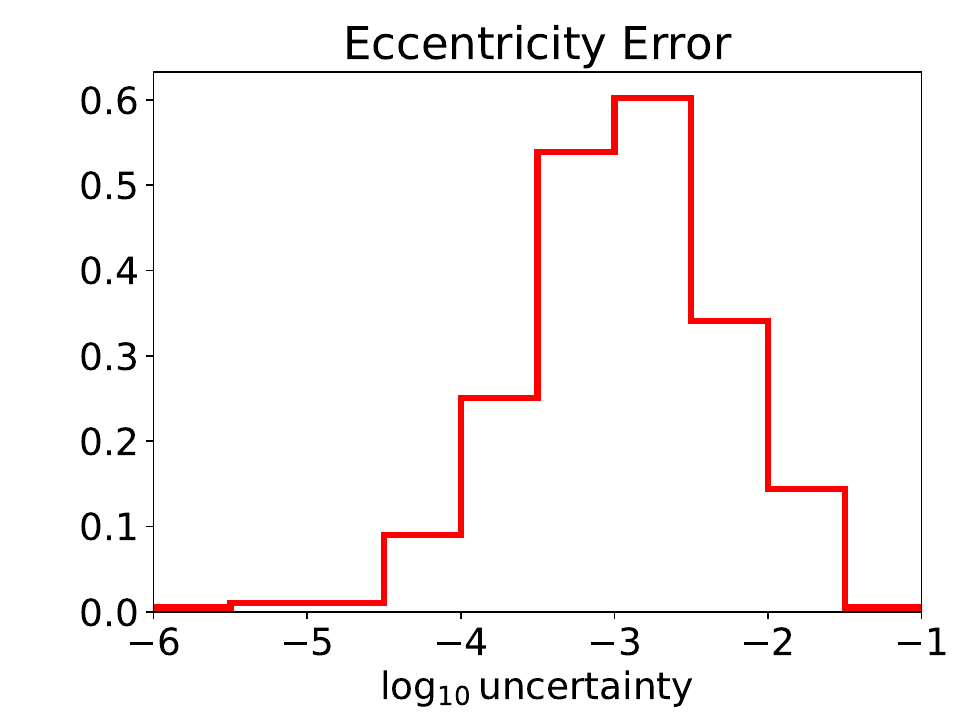}
    \caption{Distribution of uncertainty in eccentricity.}
\label{fig:appx2}
\end{figure}

We make the approximation in one of three possible ways, 
depending on the relative positions of $\mathcal{D}_e$ and $e_f^{max}=\hat{e}(e_i^{max}=0.8)$: linear for $e_i\in \left[ 0, e_i^{max} \right]$, linear in the range of $\mathcal{D}_e$, and quadratic.
The main cases are demonstrated in Figure \ref{fig:appx1}, though proportions are exaggerated for legibility.
A linear approximation is reasonable because the width of $\mathcal{D}_e$ is quite small (see Figure \ref{fig:appx2}), and thus first order Taylor expansion is sufficiently accurate.

\begin{itemize}
    \item We assume linearity across all $e_i$ when $\mathcal{D}_e$ is non-negligible at $e_f=0$ and $e_f^{max}$ is somewhere in the range of $\mathcal{D}_e$ (Figure \ref{fig:appx1}, left).
    This is, of course, the scenario where our entire range of $e_i$ falls within the range of $\mathcal{D}_e$.
    In this case, we know two points already: (0,0) and ($e_i^{max}$, $e_f^{max}$).
    We thus take $\hat{e}'$ to be $e_f^{max} / e_i^{max}$.
\end{itemize}

In all other cases, we must approximate the slope at a given location by solving for the initial eccentricity and performing a least squares fit on the last few points the solver tried.

\begin{itemize}
    \item We assume linearity across the range of $\mathcal{D}_e$ when it is negligible at both $e_f=0$ and $e_f^{max}$ and beyond (Figure \ref{fig:appx1}, middle-left).
    In this case, we do not start with information about points along the line, and so we must run additional calculations. We choose to do this at the median of $\mathcal{D}_e$. Once we have the corresponding $e_i^{median}$ and the slope, it is straightforward to construct the line equation.

    It is possible that this approximation is actually non-physical, which can happen if $\mathcal{D}_e$ is close to overlapping with (0,0) or ($e_i^{max}$, $e_f^{max}$).
    We check for this possibility by finding the value of the inverse of the approximation at either end of $\mathcal{D}_e$. If this is the case, we break it into two line segments connected at the median. 
    
    \begin{itemize}
        \item If the inverse approximation claims $e_i>0.8$ at $\mathcal{D}_{e,max}$, we fix the right end of the segment $e_i^{median} < e_i \leq 0.8$ to ($e_i^{max}$, $e_f^{max}$), because this is the largest value it could have.
        \item If it is less than zero at $\mathcal{D}_{e,min}$, we fix the left end of the segment $0 < e_i < e_i^{median}$ to (0,0), because if the system starts circularized we can expect it to end that way too.
    \end{itemize}
    
    \item The final two cases each start knowing only one of the points that must lie on our approximation. Because we must find an additional point anyway and it's not too much of an additional lift to get the slope at that point, we use a quadratic approximation.
    \begin{itemize}
        \item If $\mathcal{D}_e$ is negligible at $e_f=0$ and $e_f^{max}$ is in the range of $\mathcal{D}_e$ (Figure \ref{fig:appx1}, middle-right), we know ($e_i^{max}$, $e_f^{max}$) and find the slope and corresponding $e_i$ at $e_f^{match} = \mathcal{D}_{e,min}$.
        \item If $\mathcal{D}_e$ is not negligible at $e_f=0$ but is at $e_f^{max}$ and beyond (Figure \ref{fig:appx1}, right), we know (0, 0) and find the slope and corresponding $e_i$ at $e_f^{match} = \mathcal{D}_{e,max}$.
    \end{itemize}

    If the location of the maximum of our quadratic approximation is between $e_f^{match}$ and our known point, the approximation is non-physical: we should not switch between tidal dissipation and tidal excitement based purely on initial eccentricity.
    In this case, we model $\hat{e}(e_i)$ as a straight line between the known point and ($e_i^{match}$, $e_f^{match}$).
\end{itemize}

\bibliography{JSchuss-Prop}{}
\bibliographystyle{aasjournal}

\end{document}